\documentclass{elsart}

\usepackage{epsfig}

\newcommand{\be}{\begin{equation}}
\newcommand{\ee}{\end{equation}}
\newcommand{\bea}{\begin{eqnarray}}
\newcommand{\eea}{\end{eqnarray}}

\begin{document}

\begin{frontmatter}
\title{Transport study of charged current interactions in neutrino-nucleus reactions}
\author[unig]{W. Cassing\corauthref{cor1}}
\ead{Wolfgang.Cassing@theo.physik.uni-giessen.de}
\corauth[cor1]{corresponding author}
\author[unig]{M.~Kant}
\author[uni2]{K.~Langanke}
\author[uni3]{P.~Vogel}
\address[unig]{Institut f\"ur Theoretische Physik, %
  Universit\"at Giessen, %
  Heinrich--Buff--Ring 16, %
  D--35392 Giessen, %
  Germany}
\address[uni2]{GSI Darmstadt, %
  Planckstr. 1, %
  D--64220 Darmstadt, %
  Germany}
\address[uni3]{California Institute of Technology, %
Kellogg. Lab. 106-38, %
Pasadena, CA 91125, USA}

\begin{abstract}
  Within a dynamical transport approach we investigate
  charged current interactions in neutrino-nucleus reactions for neutrino
  energies of 0.3 - 1.5 GeV
  with particular emphasis on resonant pion production channels via the $\Delta_{33}(1232)$
  resonance. The final-state-interactions of the resonance as well
  as of the emitted pions are calculated explicitly for $^{12}C$ and
  $^{56}Fe$ nuclei and show a dominance of pion suppression at
  momenta $p_\pi >$ 0.2 GeV/c. A comparison to integrated $\pi^+$ spectra for $\nu_\mu
  + ^{12}C$ reactions with the available (preliminary) data demonstrates a
  reasonable agreement.

\end{abstract}


\end{frontmatter}

%\newpage

\section{Introduction}
The interest in neutrino induced reactions has been rapidly
increasing during the last decade while current and future long
base-line (LBL) experiments, K2K, MiniBoone, MINOS, CERN-GS, J2K
{\it etc.} aim at determining the electro-weak response of nucleons
and nuclei. In general the neutrino-nucleon/nucleus interaction consists of
the quasi-elastic scattering, resonance production and
deep inelastic scattering, the latter becoming dominant only for
neutrino energies above about 1 GeV increasing further linearly
with energy. Whereas at low neutrino energies $E_\nu <$ 0.3 GeV
the quasi-elastic channel clearly dominates, one observes an
increasing contribution from resonance excitations at higher
energies up to about 2 GeV \cite{xx1}. Here charged current (CC) reactions offer the
unique possibility to measure the $\mu^-$ in coincidence with e.g.
the charged  $\pi^{\pm}$ or  neutral $\pi^0$ in order to
extract information on the resonance transition form factors since the
4-momentum of the lepton fixes the transferred energy $\nu$ and
4-momentum squared $q^2$.

A variety of models (and event generators) treat neutrino nucleus
reactions \cite{P1,P2,P3,NUANCE,NEUT,NEUGEN,NUX-FLUCA} and
incorporate selected aspects of scattering on nuclei like
Pauli-blocking, nuclear Fermi motion, nucleon binding energies, pion
rescattering and charge exchange reactions. However, most of the
models lack an explicit propagation of the resonances and their
in-medium decay to pions and nucleons as well as more realistic
coordinate (and momentum) distributions of the target nucleons. In
order to study the importance of these effects we employ in this
work a transport approach that has been tested in a wide range of
$\pi, p$ reactions on nuclei as well as relativistic nucleus-nucleus
collisions. For reviews we refer the reader to Refs.
\cite{Cass90,Cass99}. This transport model -- or related versions --
have been also used and tested in the description of the
photo-nuclear and electron induced reactions with the momentum and
energy transfer comparable to those considered here \cite{Effe,Mueh}
as well as at much higher energies \cite{Falter}. This is
particularly relevant, since the nuclear response to electromagnetic
probes is quite analogous to the weak probes addressed in this
study.

For the description of the elementary process on nucleons we use the
model of Rein and Sehgal \cite{Segal} - with slight modifications -
that has been often employed for an estimate of neutrino cross
sections for resonance transitions. It is based on quark oscillator
wave functions to evaluate the form factors as developed by Feynman
et al. \cite{Feynman}. Alternatively, one may also parametrize the
neutrino-nucleon-resonance vertex by introducing 'proper'  vector
and axial-vector form factors. The differential cross sections are
well known \cite{r5} and usually given in terms of helicity
amplitudes. Our present investigation is focused on
final-state-interactions (FSI) of the excited $\Delta(1232)$
resonance and the emitted pion. Total $\pi^+$ cross sections from
$^{12}C$ targets will be presented in comparison to preliminary data
as a function of the neutrino energy $E_\nu$ from 0.3 to 1.5 GeV. We
also provide energy differential $\pi^\pm, \pi^0$ spectra for
$^{12}C$ and $^{56}Fe$ at $E_\nu$ = 1 GeV and perform a differential
analysis of final-state-interactions (FSI) on both targets that are
used in the MiniBoone and MINOS detectors as target material. Note
that alternative forms of the neutrino-nucleon-resonance vertex will
modify the double differential cross sections but have no severe
impact on the strength of the FSI.

\section{Neutrino-nucleon  reactions}
We consider the interactions of a $\nu_{\mu}$ neutrino with a
nucleon that lead to a muon and pion in the final state, i.e. the charged
current (CC) interactions
\be
\nu_{\mu} + p \rightarrow \mu^- + \pi^+ +p
\ee
\be
\nu_{\mu} + n \rightarrow \mu^- + \pi^+ +n
\ee
\be
\nu_{\mu} + n \rightarrow \mu^- + \pi^0 +p .
\ee
Neutral current (NC) interactions are
\be
\nu_{\mu} + p \rightarrow \nu_{\mu}  + \pi^0 +p
\ee
\be
\nu_{\mu} + p \rightarrow \nu_{\mu}  + \pi^+ +n
\ee
\be
\nu_{\mu} + n \rightarrow \nu_{\mu}  + \pi^- +p
\ee
\be
\nu_{\mu} + n \rightarrow \nu_{\mu}  + \pi^0 +n .
\ee
The NC reactions with the $\pi^0$ production, eqs. (4) and (7), represent a dangerous background
in the long-baseline neutrino oscillation studies, since they could be confused with the
quasi-elastic electron production caused by neutrino-oscillations. We will consider the
NC pion production separately in future work.

We denote the energy transfer by $\nu = E_\nu - E_l$, where $E_l$
is the energy of the final lepton, the 4-momentum transfer squared
by $q^2= \nu^2-Q^2 = M^2-m_n^2 - 2 m_n \nu$ with $M, m_n, Q$ denoting
the resonance mass, nucleon mass and 3-momentum transfer, respectively. We then
obtain for the invariant energy (with the nucleon at rest) $W =
\nu+m_n$ or alternatively $W^2=M^2 + Q^2$. The double differential
cross section for neutrino $\Delta$-resonance excitation then reads
\be \label{cross}
\frac{d\sigma}{dq^2 d \nu} = - \frac{(G_F \cos(\Theta_C))^2}{4
\pi^2} \frac{q^2}{Q^2} \kappa \ [u^2 \sigma_L + v^2 \sigma_R + 2u v
\sigma_s]
\ee
with
\be u= \frac{E_\nu + E_l+Q}{3 E_\nu}, \ v= \frac{E_\nu + E_l-Q}{3
E_\nu},\ \kappa= \frac{M^2-m_n^2}{2 m_n}.
\ee
The $\sigma$ terms read
\be
\label{E1}
\sigma_{L,R}(q^2,W) =  \frac{ W M}{ \kappa m_n} \sum_{j_z}
|\langle N,j_z\mp1|F_{\mp}|\Delta,j_z \rangle|^2 \ 2 W A(W,M) ,
\ee
\be
\label{E2}
\sigma_{s}(q^2,W) = -\frac{ W m_n}{\kappa M} \frac{Q^2}{q^2} \sum_{j_z}
|\langle N,j_z\mp1|F_{0}|\Delta,j_z \rangle|^2 \ 2 W A(W,M) ,
\ee
where $A(W,M)$ denotes the $\Delta$ spectral function - to be
specified below - which replaces the conventional term $2 \pi  W \delta(W^2-M^2)$.
The $N-\Delta$ transition amplitudes $F_{\mp}, F_{0}$ in (\ref{E1}, \ref{E2}) are well
known in principle \cite{Segal}, however, involve rather uncertain
 vector and axial-vector form factors which we adopt in the form
 \be \label{form}
 G^{V,A}(q^2) = \sqrt{1 - \frac{q^2}{4 m_n^2}}\  \left( 1 -
 \frac{q^2}{M^2_{V,A}} \right)^{-2}
 \ee
 with $M_{V,A}$ specifying  characteristic mass (or inverse length) scales. We
 use the  parameters \cite{Kolbe} $M_V = 0.843$ GeV and $M_A= 1.032$ GeV in the following
 calculations (default parameter set $A$).

The $\Delta$ spectral function is taken in the form (with $M_\Delta
 = 1.232$ GeV)
\be \label{spectral} A(W,M_\Delta) = \frac{W
\Gamma(W)}{(W^2-M_\Delta^2)^2 + W^2 \Gamma^2(W)}, \ee where the
width $\Gamma$ explicitly depends on $W$. In vacuum a good
parametrization is given by \cite{Wolf90} \be \Gamma(W)= \Gamma_0
\left(\frac{q}{q_r} \right)^3 \frac{M_\Delta}{W} \left(
\frac{v(q)}{v(q_r)} \right)^2 , \hspace{1cm} v(q) =
\frac{\beta^2}{\beta^2+q^2}, \ee with \be  q^2(W) =
(W^2-(m_n+m_\pi)^2)(W-(m_n-m_\pi)^2)/(4 W^2) . \ee Following
\cite{Wolf90} the parameters are $\Gamma_0 = $ 110 MeV, $\beta$ =
0.3 GeV. Furthermore, $q_r = q(M_{\Delta})$ while $m_\pi$ denotes
the pion mass. We have checked that alternative parametrizations
only moderately change the shape of the cross section (\ref{cross})
in line with Ref. \cite{Tina_dip}.

The results for ${d\sigma}/{dq^2 d \nu}$ for the channel $\nu_\mu +
p \rightarrow \mu^- + \Delta^{++}$ are displayed in Fig. 1 for a
neutrino energy $E_\nu$ = 1 GeV and show a characteristic bump in
the transferred energy $\nu$ due to the spectral function
(\ref{spectral}) which softens and kinematically shifts
with increasing negative $q^2$. A
similar functional dependence on $\nu$ and $q^2$ is known from $e^-
+ A$ reactions \cite{Mueh}.

%%%%%%%%%%%%%%%%%%%%%%%%%%%%%%%%%%%%%%%%%%%%%%%%%%%%%%%%% Fig.1%%%%%%%%
\begin{figure}[htb!]
  \begin{center}
    \includegraphics[width=9.5cm,angle=-90]{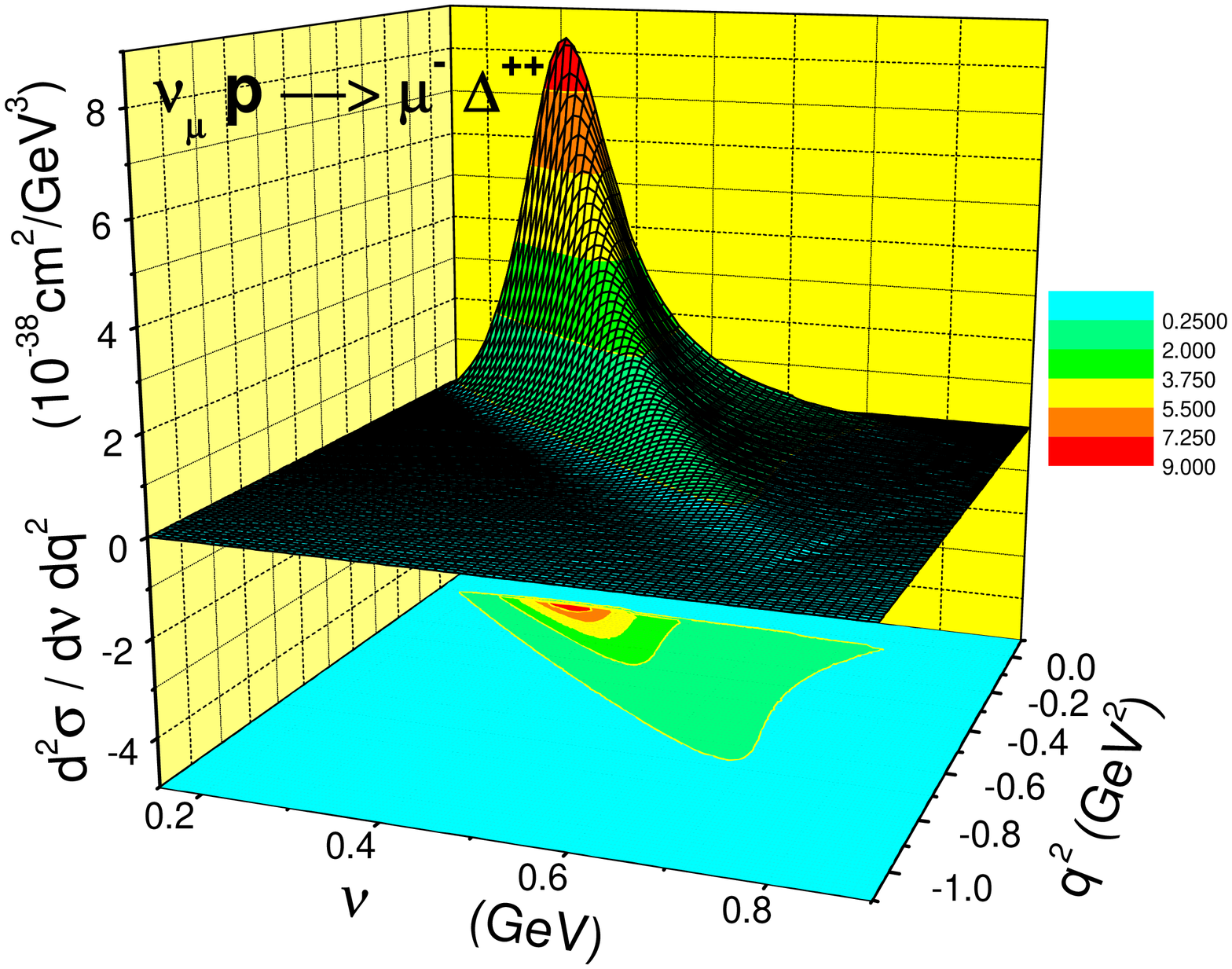}
    \caption{
      The double differential cross section $d\sigma/d \nu d q^2$ for the reaction
      $\nu + p \rightarrow \mu^- + \Delta^{++}$ as a function of the transferred energy $\nu$
      and the 4-momentum transfer squared $q^2$ for a neutrino energy $E_\nu$ = 1 GeV.}
    \label{fig1}
  \end{center}
\end{figure}
%%%%%%%%%%%%%%%%%%%%%%%%%%%%%%%%%%%%%%%%%%%%%%%%%%%%%%%%%%%%%%%%%%%%%%%
Since presently no double differential data are available for a
detailed comparison and fixing of the form factors (\ref{form}), we
use cross sections that are integrated over $\nu$ and $q^2$. Note
that the data in Fig. 2 \cite{xxp} correspond to reactions on $H_2$
and $D_2$ and differ significantly from each other. Our standard
'historical' parameter  set $A$ \cite{Kolbe} falls slightly low for
both reactions, while calculations with a modified parameter set $B$
($M_V=$ 1 GeV, $M_A = $ 1.2 GeV) perform better (though not
perfectly). Our model gives a ratio of the $\pi^+$ cross section
from the proton relative to the neutron of 3 which is determined by
a relative Clebsch-Gordon coefficient of $\sqrt{3}$ in the decay
amplitude. The original Rein-Sehgal model \cite{Segal} has an
additional factor $\sqrt{3}$ in the production amplitude for the
$\Delta^{++}$ relative to the $\Delta^+$, which gives a total ratio
of 1/9 for the cross sections $\nu_\mu + n \rightarrow \mu^- + \pi^+
+ n$ relative to $\nu_\mu + p \rightarrow \mu^- + \pi^+ + p$. Since
the data in Fig. 2 are not compatible with a ratio 1/9 but favor a
ratio 1/3 we have 'redefined' the production matrix element to be
identical for $\Delta^{++}$ and $\Delta^+$. Note that also higher
baryon resonances may contribute to $\pi^+$ production - especially
for the $\nu_\mu + n \rightarrow \mu^- + \pi^+ + n$ channel - which
are discarded in this study. These resonances have independent
transition form factors that presently are not controlled by data
and require further parameters.  The explicit dependence of the
cross section on $E_\nu$ reflects both the form factors employed as
well as the detailed form of the spectral function (\ref{spectral}).

In view of the large uncertainties in the data we do not aim at fitting the form
factors here and consider instead  reactions on nuclei by employing a coupled-channel
transport model.

%%%%%%%%%%%%%%%%%%%%%%%%%%%%%%%%%%%%%%%%%%%%%%%%%%%%%%%%% Fig.1%%%%%%%%
\begin{figure}[htb!]
  \begin{center}
    \includegraphics[width=8cm,angle=-90]{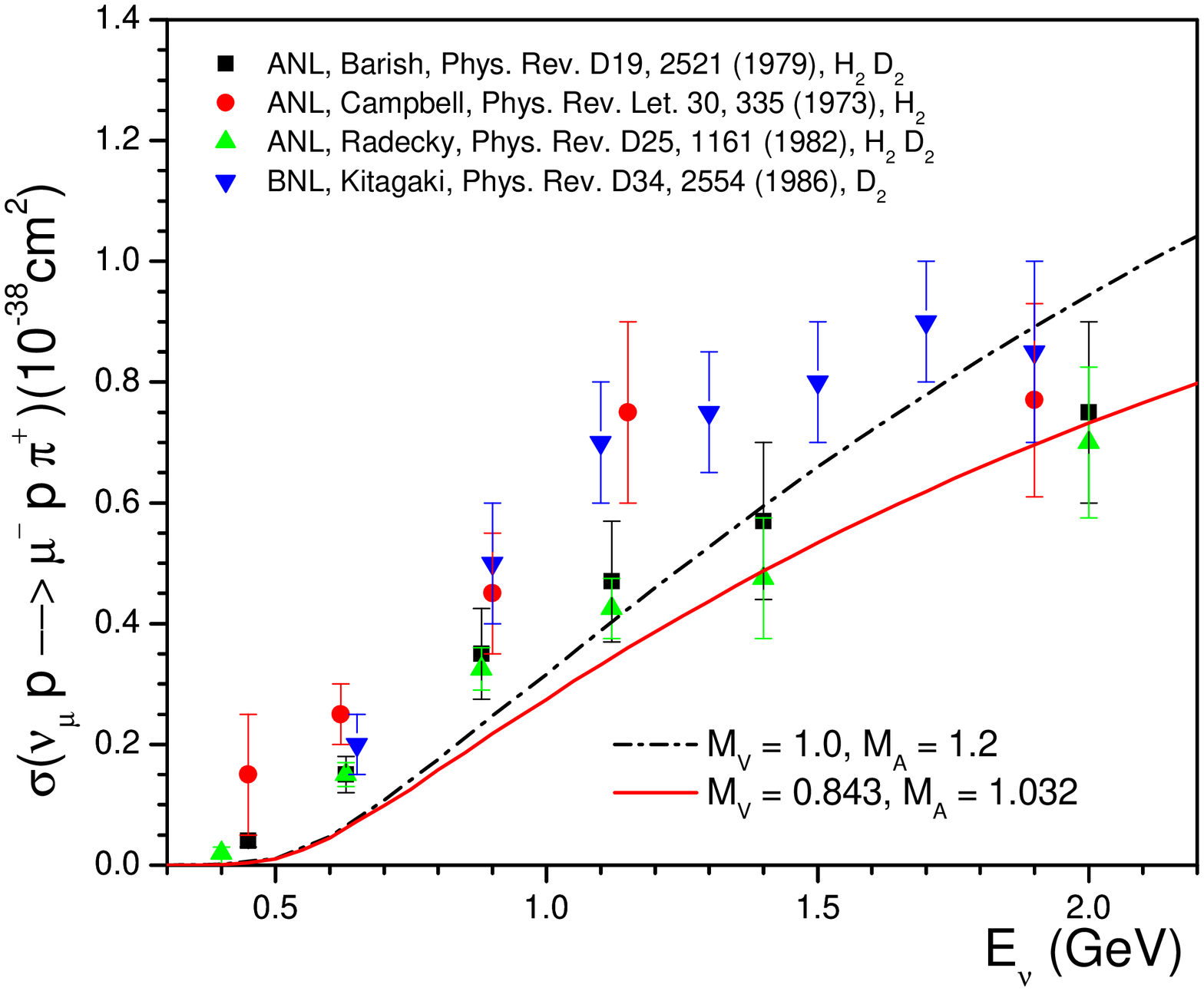}

    \vspace{-1.2cm}
    \includegraphics[width=8cm,angle=-90]{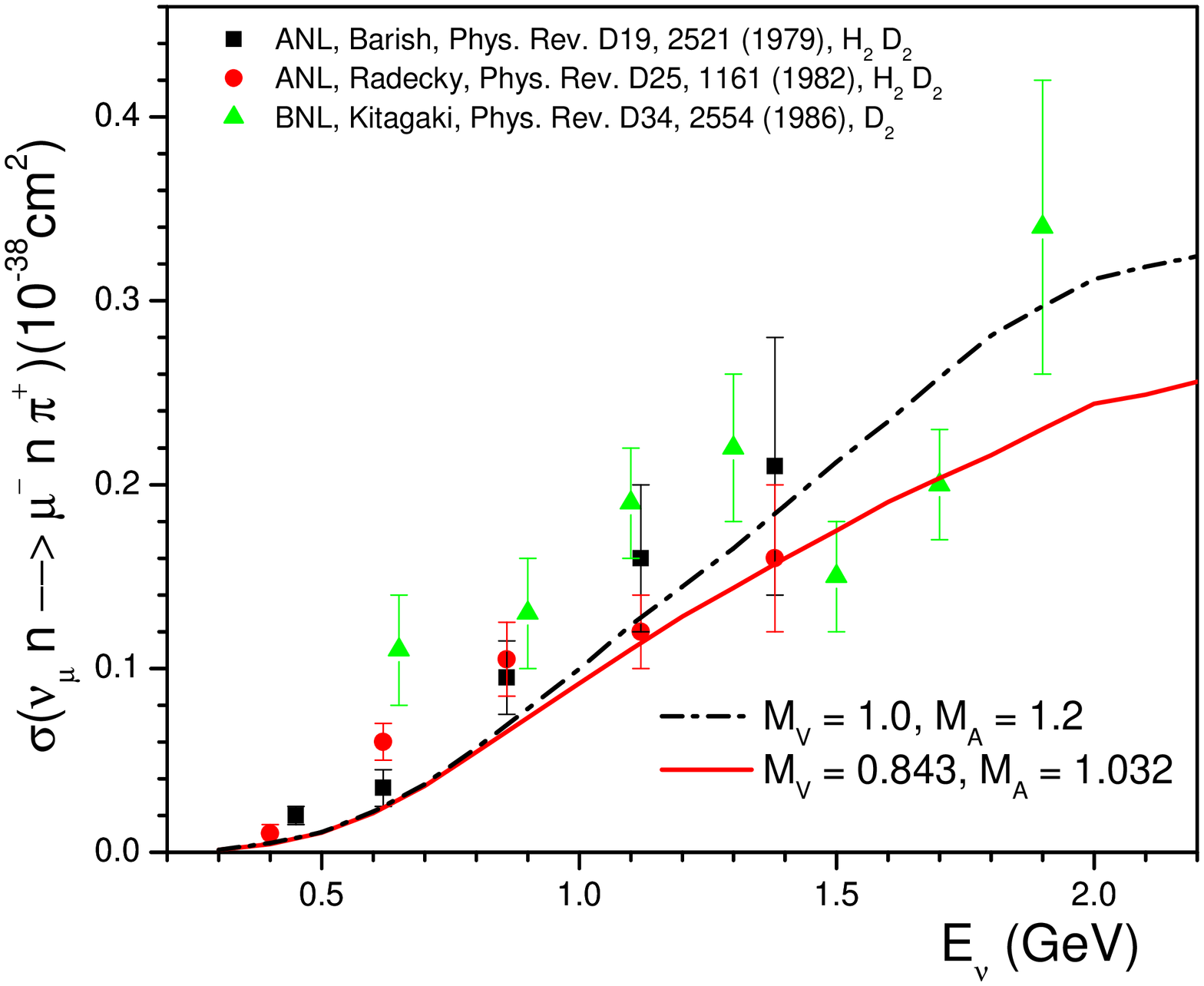}
    \caption{
      The calculated cross sections for the reactions
      $\nu + p \rightarrow \mu^- + \pi^{+}+p$ (upper part) and
      $\nu + n \rightarrow \mu^- + \pi^{+}+n$ (lower part) as a function of the
       neutrino energy $E_\nu$ for the standard parameter set $A$ ($M_V= 0.843$ GeV, $M_A =
       1.032$ GeV) and the modified set $B$ ($M_V=$ 1 GeV, $M_A = $ 1.2 GeV). The data
       are taken from  Refs. \cite{xxp}, respectively.}
    \label{fig01}
  \end{center}
\end{figure}
%%%%%%%%%%%%%%%%%%%%%%%%%%%%%%%%%%%%%%%%%%%%%%%%%%%%%%%%%%%%%%%%%%%%%%%

\section{Neutrino nucleus reactions}

Relativistic transport models have been originally developed for
nucleus-nucleus collisions starting from the Fermi-energy domain up
to ultra-relativistic energies ($\sqrt{s} \approx$ 200 GeV) in order
to extract physical information from nucleus-nucleus collisions
relative to scaled proton-nucleus reactions
\cite{Cass90,Cass99,URQMD1,URQMD2,Ehehalt,excita}. These models
incorporate a variety of in-medium effects such as mean-field
potentials - i.e. the real part of the retarded selfenergy - as well
as finite life time effects of hadrons due to multiple interactions
(decays) in the medium. The latter are encoded in dynamical spectral
functions for the hadrons of interest that are fully specified in
terms of the real and imaginary parts of the hadron selfenergy. We
recall that an off-shell transport version - suitable for the
propagation of particles with finite spectral width - is available,
too \cite{Juchem}. While the initial studies had been devoted to
$A+A$ collisions, these models have been used in the description of
pion-nucleus \cite{Golubeva} and photo-nuclear reactions
\cite{Effe,Mueh,Mesh} in more recent years as well. The advantage of
such transport models is that they are suited for a wide variety of
reactions and thus can relate e.g. photo-nuclear reactions to
pion/proton induced reactions as well as neutrino-nucleus reactions
(as addressed here).

In this work we employ the CBUU\footnote{Coupled-Channel
Boltzmann-Uehling-Uhlenbeck} approach~\cite{Wolf} which is based on
a  set of  transport equations for the phase-space distributions
$f_{h} (x,p)$ of hadron $h$ that are coupled via collision terms
describing the mutual interaction (or decay and formation) rates in phase space. The
mean-fields for the baryons  are the same as in Ref.
\cite{excita} while the description of hadronic collisions is
identical to Ref. \cite{Cass99}.

For the neutrino-nucleus reactions we employ a perturbative
treatment  as in the case of photo-nuclear reactions \cite{Falter}
which allows us to obtain good statistics even for weakly
interacting probes. In this respect the total neutrino-nucleus
reaction is divided into two parts: i) the initial neutrino
interaction on a single nucleon of the target (including its local
Fermi motion and mean-field potential) and ii) the propagation and
decay of the excited resonance (as well as the decay products) in
the residual nucleus. Each excited resonance in step i) is assigned
a weight $W_i$ proportional to the double differential cross section
$d \sigma/d \nu d q^2$ where $\nu$ and $q^2$ are selected by Monte
Carlo in the kinematical allowed regime according to the
differential distributions displayed e.g. in Fig. 1 for $E_\nu$ = 1
GeV. A large number of simulations ($\sim 10^4$) are performed such
that the sum over the weights $W_i$ - normalized by the number of
events - is equal to the $\nu$ and $q^2$ integrated cross section per nucleon.
Note that for fixed $\nu$ and $q^2$ and each individual 4-momentum
$q_N$ of the nucleon the invariant energy as well as the 3-momentum
of the excited resonance are fully determined by the energy and
momentum conservation.

We then follow the motion of the perturbative hadrons (here
$\Delta$'s, decay pions and nucleons) within the full background of
residual nucleons by propagating the $\Delta$'s with 2/3 of the
nucleon potential - neglecting pion potentials in this study - and
compute their collisional history with the residual nucleons of the
target. A reduced $\Delta$ potential - relative to the nucleon potential -
is adopted in order to improve
the description of pion spectra from p+A reactions and to
incorporate the findings from Peters {\it et al.} \cite{Peters} for
coherent photo-production of pions on nuclei. In order to trace back
the FSI of the hadrons the $\Delta$-production vertices are recorded
with their individual weight (including their 4-momenta) as well as
the 4-momenta of the final pions.

We point out that an alternative transport description of
neutrino-nucleus reactions on the basis of the GiBUU transport model
has been initiated by T. Leitner {\it et al.} in Refs. \cite{Tina}.

\begin{figure}[htb!]
  \begin{center}
    \includegraphics[width=8cm,angle=-90]{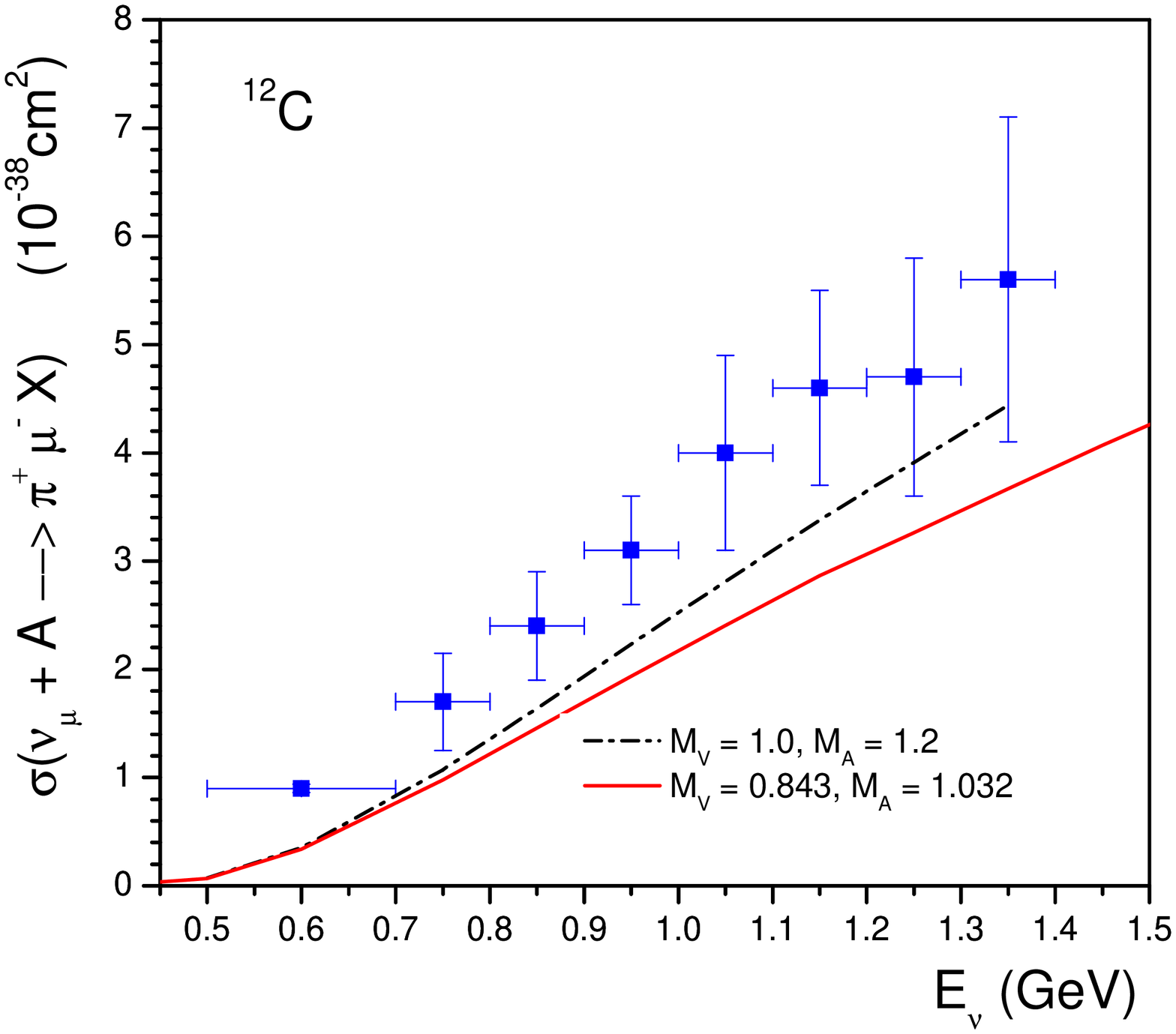}
    \caption{
      The total cross section $\sigma_{\pi^+}$ for the
      reaction  $\nu + ^{12}C  \rightarrow \mu^- + \pi^{+}+X$ as a function of the
      neutrino energy $E_\nu$ for the two parameter sets discussed in the text.
      The preliminary data have been taken from Ref. \cite{xxn}.}
    \label{fig3c}
  \end{center}
\end{figure}

Here we present the calculated cross sections for $\pi^+$ meson
production in CC reactions on $^{12}C$ targets. The results are
displayed in Fig. \ref{fig3c} for the two parameter sets discussed
in the context of the elementary cross sections shown in Fig. 2. We
find that the parameter set $A$  underestimates the preliminary data
on $^{12}C$ from Ref. \cite{xxn} as in case of the proton or neutron
target in Fig. 2 while the parameter set $B$ performs better (as
expected from Fig. 2). Note, however, that any
agreement/disagreement is primarily due to the 'uncertain'
axial-vector form factor $G^A(q^2)$ such that no final conclusions
can be drawn from the comparison in Fig. \ref{fig3c}.

%%%%%%%%%%%%%%%%%%%%%%%%%%%%%%%%%%%%%%%%%%%%%%%%%%%%%%%%% Fig.1%%%%%%%%
%%%%%%%%%%%%%%%%%%%%%%%%%%%%%%%%%%%%%%%%%%%%%%%%%%%%%%%%%%%%%%%%%%%%%%%

\begin{figure}[htb!]
  \begin{center}
    \includegraphics[width=7.3cm,angle=-90]{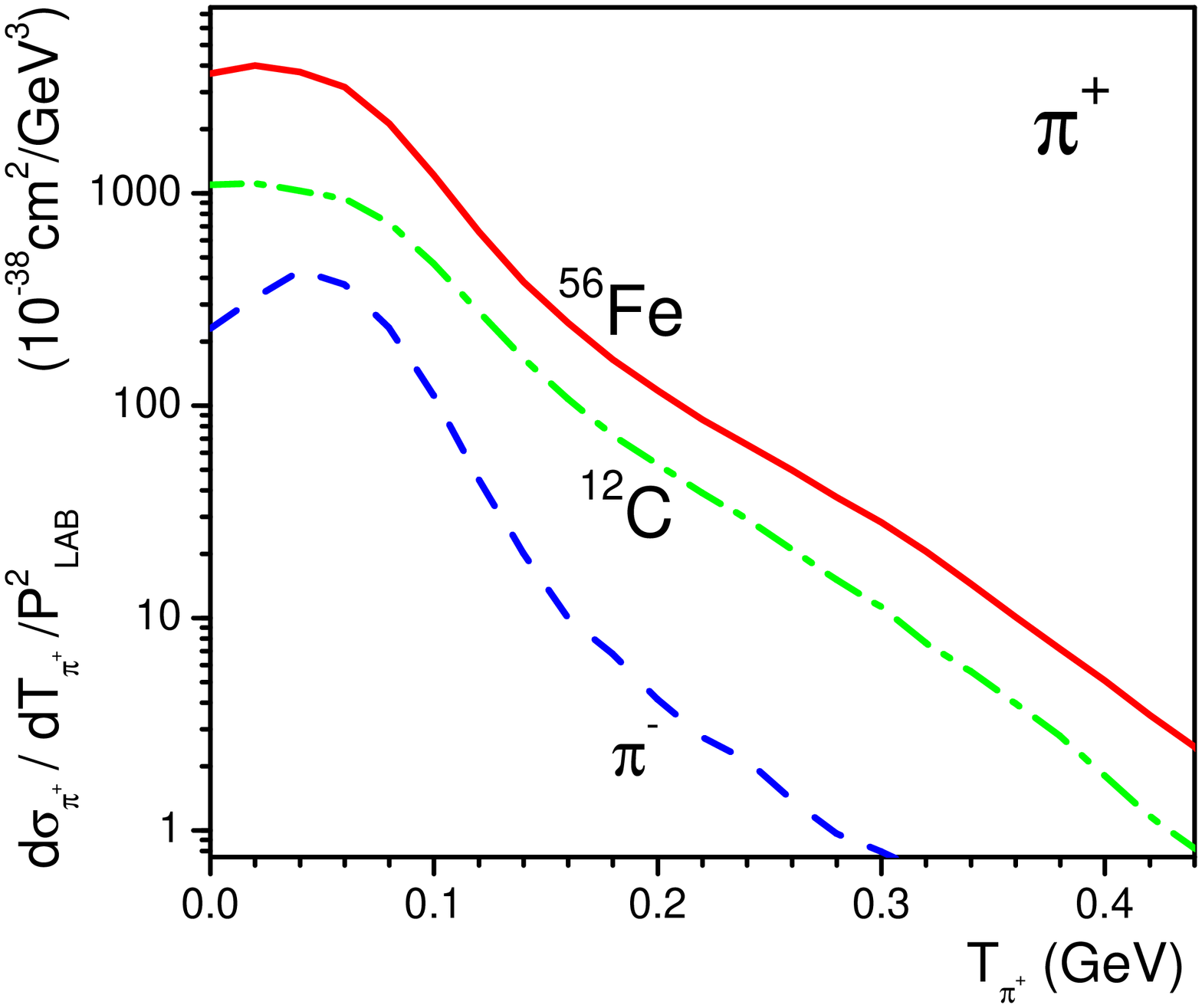}

    \vspace{-1.0cm}
    \includegraphics[width=7.3cm,angle=-90]{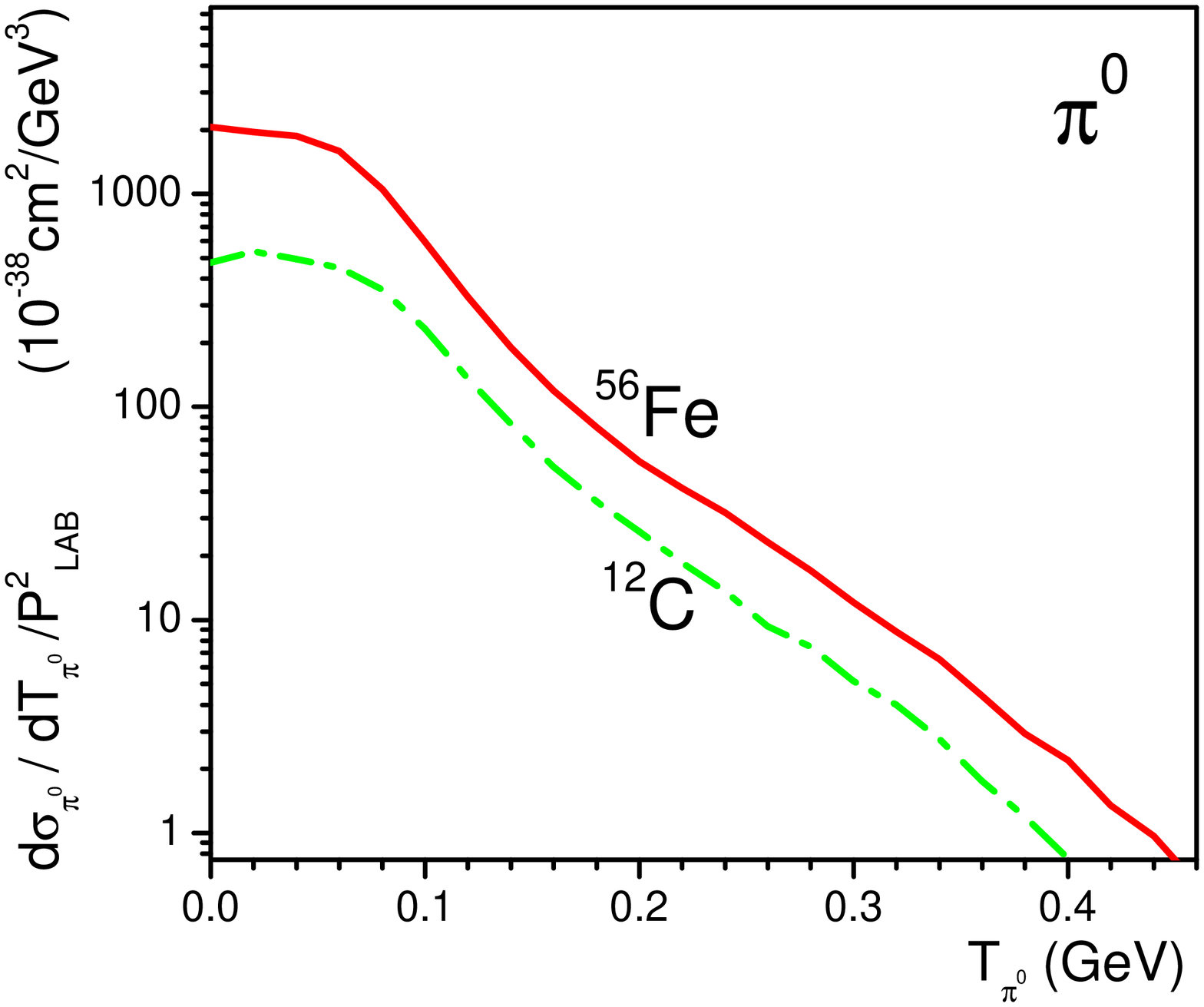}
    \caption{
      The differential cross sections $d\sigma_{\pi^+}/d T_\pi/P_{lab}^2$ (upper part) and
      $d\sigma_{\pi^0}/d T_\pi /P_{lab}^2$ (lower part) for the reactions
      $\nu + ^{12}C ((^{56}Fe) \rightarrow \mu^- + \pi +X$ as a function of the
      pion kinetic energy $T_\pi$ in the laboratory
      for a neutrino energy $E_\nu$ = 1 GeV. The $\pi^-$
      differential spectrum (in case of the $^{56}Fe$ target) is shown by
      the dashed line (in the upper part) and is entirely due to
      final state interactions in the target nucleus.}
    \label{fig3}
  \end{center}
\end{figure}

As an example the  differential pion ($\pi^+, \pi^0$) spectra in the
pion kinetic energy $T_\pi$ - divided by the pion laboratory
momentum $P_{lab}$ squared - are shown in Fig. \ref{fig3} for a
$^{12}C$ and $^{56}Fe$ target at the neutrino energy $E_\nu$ = 1
GeV. The differential spectra $d\sigma_\pi/dT_\pi/P_{lab}^2$ for
$^{12}C$ and $^{56}Fe$ show a shoulder at low $T_\pi$ and are
roughly exponential for high pion energies. The shape of the spectra
is similar for both targets but they do not simply scale by a factor
of 36/8 = 4.5 as expected from a scaling of the elementary cross
sections. (This factor comes about as follows: (26+30/3)/(6+6/3)
when counting the protons and neutrons with their excitation to
$\Delta^{++}$ and $\Delta^+$, respectively, and  the relative decay
to $\pi^+$. For $\pi^0$ the decay fraction of the $\Delta^+$ is 2/3,
however, no direct $\pi^0$'s can be produced in CC reactions on
protons. The corresponding scaling factor thus is 30/6= 5.) In case
of the $Fe$ target we also display the $\pi^-$ spectra (upper part)
that are down by an order of magnitude relative to the $\pi^+$
spectra and stronger localized at low pion energies. Note that no
direct negative pions can be produced in the CC neutrino reactions
on nucleons and therefore the final $\pi^-$ (or $\pi^-/\pi^+$ ratio)
provide a sensitive probe for the strength of the FSI.

%%%%%%%%%%%%%%%%%%%%%%%%%%%%%%%%%%%%%%%%%%%%%%%%%%%%%%%%% Fig.1%%%%%%%%
%%%%%%%%%%%%%%%%%%%%%%%%%%%%%%%%%%%%%%%%%%%%%%%%%%%%%%%%%%%%%%%%%%%%%%%

Since the differential pion spectra suffer from the same
uncertainties as the integrated spectra we turn to ratios of cross
sections that reflect dominantly the influence of the FSI. For this
purpose we consider the ratios \be \label{ratio} R_x({\bf p}) =
\frac{(d^3 \sigma_x/d^3p)_{fin}}{(d^3 \sigma_x/d^3p)_{in}} \ee where
the initial differential spectra in the denominator of Eq.
(\ref{ratio}) are determined from the decay of the primarily
produced $\Delta$ resonances without FSI. Fig. 5 shows the ratio
$R_0$ of the final differential $\pi^0$ momentum spectrum to the
neutral pion spectrum without FSI for $\nu_\mu + ^{56}Fe$ at $E_\nu$
= 1 GeV as a function of the longitudinal momentum $P_z$ and
transverse momentum $P_T$ (with respect to the direction of the
incoming neutrino). This ratio is roughly constant ($R \sim 0.4$)
for $\sqrt{P_T^2+P_z^2} >$ 0.2 GeV/c and exceeds unity for lower
pion momenta in the laboratory due to strong FSI of the $\Delta$'s
and decaying pions with nucleons. We recall that $N\Delta$
interactions in the medium involve - apart from energy and momentum
exchange -- charge transfer as well as $\Delta N \rightarrow NN$
absorption reactions. The latter cross sections are obtained from
refined detailed balance relations that incorporate the finite (mass
dependent) width of the $\Delta$-resonance \cite{Wolf}. The strong
reduction of the pion spectrum at higher pion momenta is essentially
due to $\pi + N \rightarrow \Delta$ resonant scattering.

\vspace{0.5cm}
%%%%%%%%%%%%%%%%%%%%%%%%%%%%%%%%%%%%%%%%%%%%%%%%%%%%%%%%% Fig.1%%%%%%%%
\begin{figure}[htb!]
  \begin{center}
    \includegraphics[width=8cm,angle=-90]{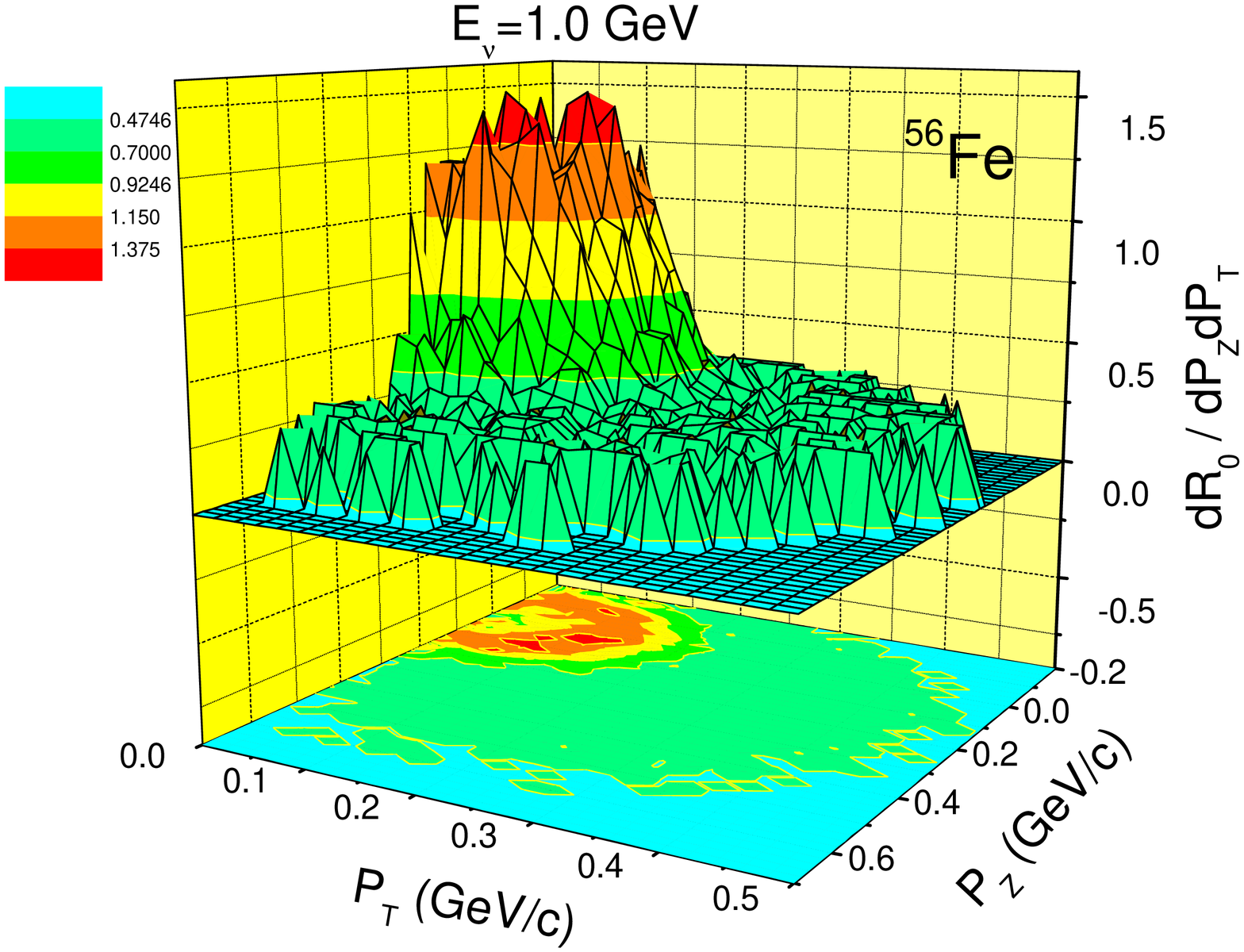}
    \caption{
      The ratio $R_0$ of the final differential $\pi^0$ momentum spectrum to the $\pi^0$
      spectrum without FSI for $\nu_\mu + ^{56}Fe$ at $E_\nu$ = 1 GeV as a function of the
      longitudinal momentum $P_z$ and transverse momentum $P_T$ (with respect to the direction
      of the incoming neutrino).}
    \label{fig3d}
  \end{center}
\end{figure}
%%%%%%%%%%%%%%%%%%%%%%%%%%%%%%%%%%%%%%%%%%%%%%%%%%%%%%%%%%%%%%%%%%%%%%%

Apart from the overview in Fig. 5 it is, furthermore, of interest to
quantify the strength of the FSI e.g. as a function of the pion
kinetic energy in the laboratory $T_\pi$. All ratios displayed in
Fig. 6 for $\pi^+$ (upper part) and $\pi^0$ (lower part) show a
characteristic dip at low kinetic energy $T_\pi$ followed by a bump
(with a ratio above unity due to pion rescattering) and an
approximate plateau for $T_\pi >$ 0.15 GeV which is lower for the
heavier $Fe$-target than for the light $C$-nucleus. These
differential FSI effects can also be directly assessed by taking
experimental differential ratios of spectra for heavy and light
targets.

Note that charged current reactions with pion production and
subsequent absorption of the pion might be misinterpreted
experimentally as the quasi-elastic CC  reactions and lead to an
overestimation of the CC quasi-elastic cross sections. In case of
the $^{12}C$ target we find an integrated $\pi^+$ ($\pi^0$)
absorption of 12\% (14\%). When imposing a detection threshold for
pions of $T_\pi$= 100 MeV in the laboratory the relative absorption
amounts to 33\% (35\%) for $\pi^+$ ($\pi^0$). For the $^{56}Fe$
target the respective numbers are 35\% (42\%) for the integrated
spectra and 58\% (64\%) for an energy cut of $T_\pi$=100 MeV.

%%%%%%%%%%%%%%%%%%%%%%%%%%%%%%%%%%%%%%%%%%%%%%%%%%%%%%%%% Fig.1%%%%%%%%
\begin{figure}[htb!]
  \begin{center}
    \includegraphics[width=6.5cm,angle=-90]{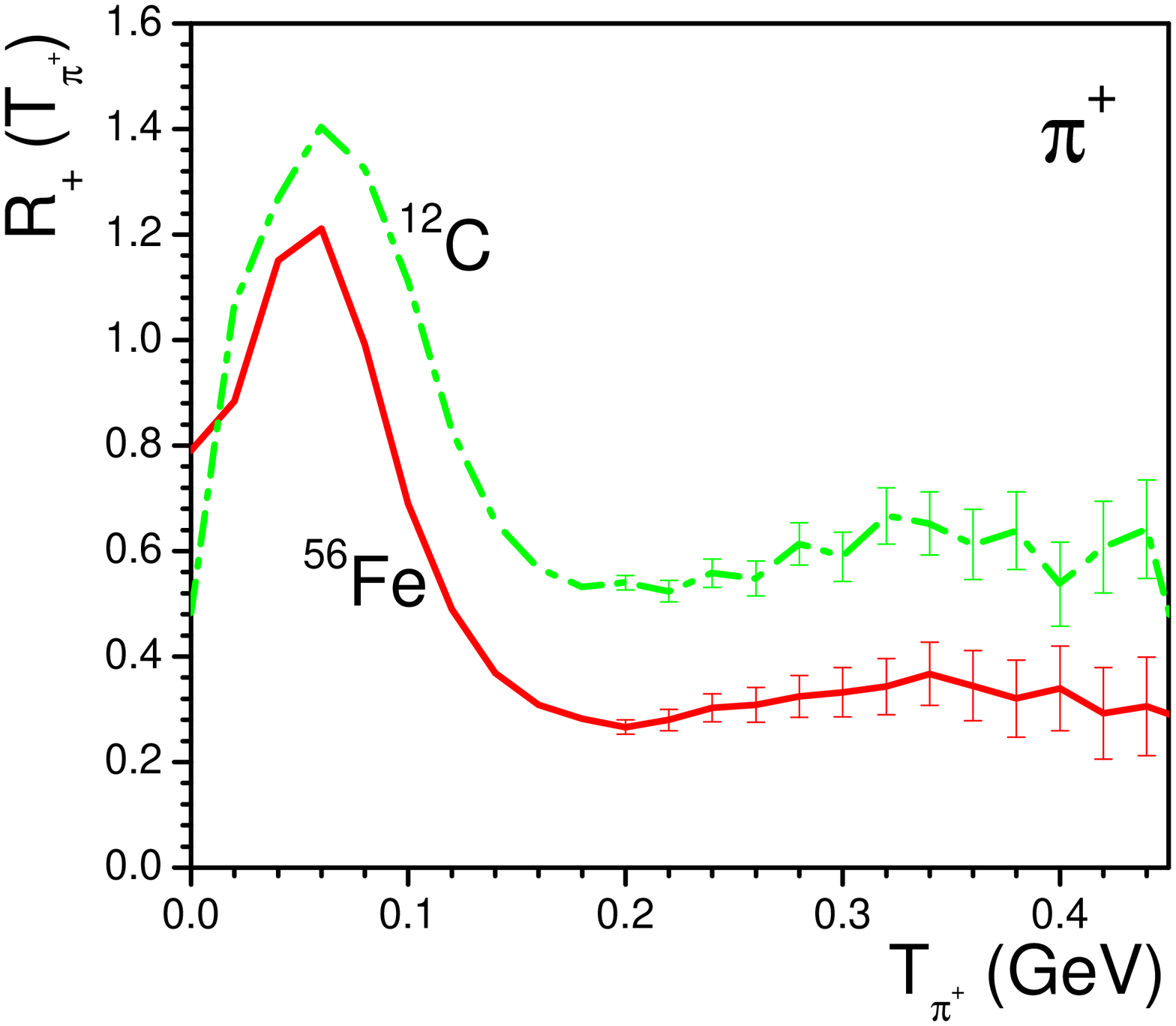}

    \vspace{-0.7cm}
    \includegraphics[width=6.5cm,angle=-90]{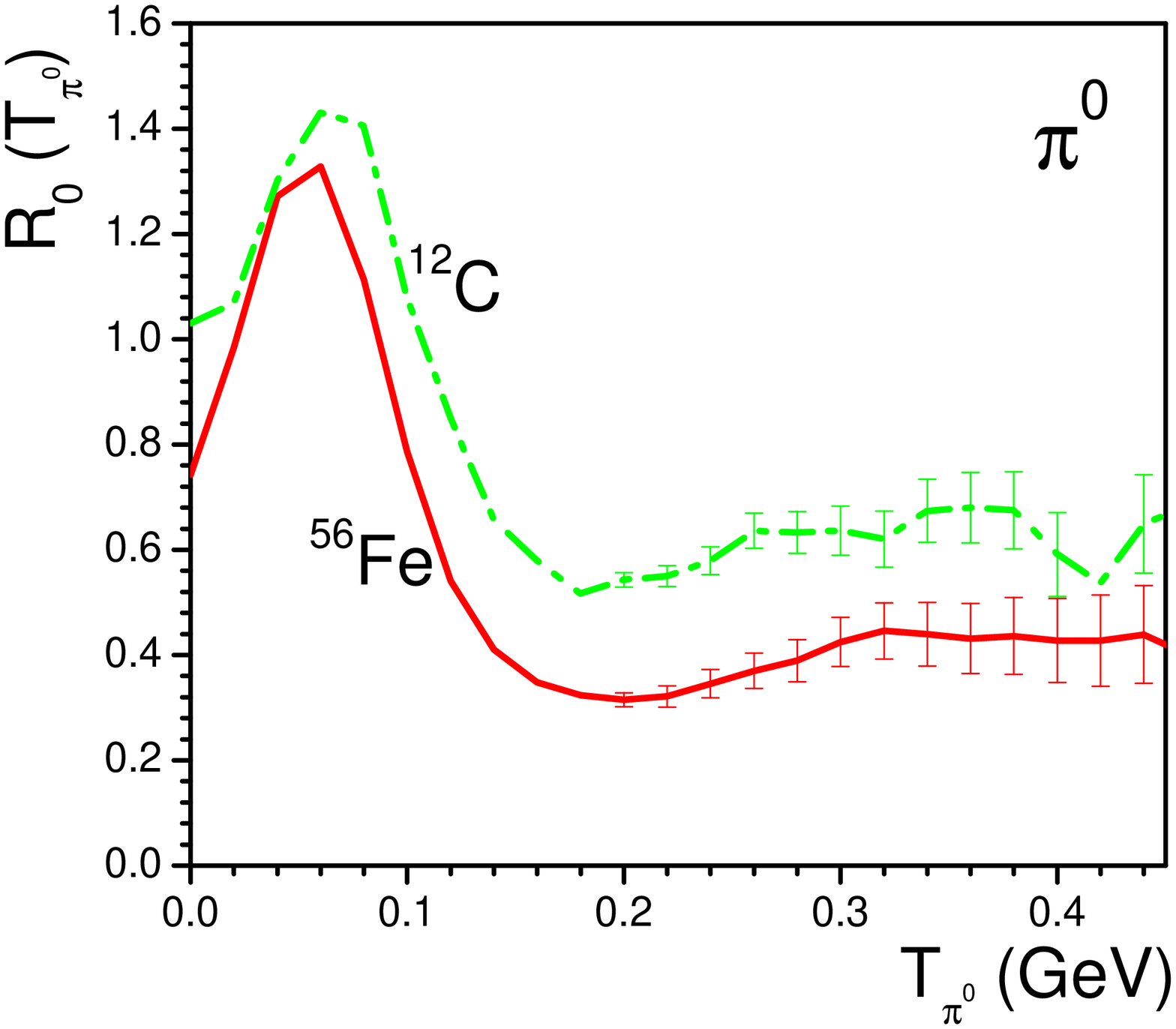}
    \caption{
      The differential ratios $R_+(T_\pi)$ (upper part) and $R_0(T_\pi)$ (lower part)
      for the  reactions
      $\nu_\mu + ^{12}C (^{56}Fe) \rightarrow \mu^- + \pi^{+}(\pi^0)+X$ as a function of the
      pion kinetic energy $T_\pi$ in the laboratory for a neutrino energy $E_\nu$ = 1 GeV.
      The error bars
      denote the statistical uncertainty in determining the ratios numerically
      due to the low cross sections at large pion energies (cf. Fig. 4).}
    \label{fig5}
  \end{center}
\end{figure}
%%%%%%%%%%%%%%%%%%%%%%%%%%%%%%%%%%%%%%%%%%%%%%%%%%%%%%%%%%%%%%%%%%%%%%%

\section{Summary}
In this study we have reported first dynamical transport
calculations for neutrino-nucleus reactions in the energy range from
0.3 - 1.5 GeV. The advantage of our transport description is that it
is suited also for a wide variety of reactions and thus can relate
neutrino-nucleus reactions to photo-nuclear reactions or pion/proton
induced reactions on nuclei. Furthermore, standard in-medium effects
such as mean-field potentials or in-medium broadening of hadronic
resonances can be exploited within the same framework. Effects of
Fermi motion and Pauli-blocking of nucleons are incorporated by
default.

The method employed here is quite general and can be applied to a
wide variety of nuclei and reaction channels that incorporate all
possible final state interactions. Furthermore, the transport
approach can be used for an event-by-event analysis and correlate
the momenta of different particles in each event, e.g. emitted
nucleons with pions {\em etc.}. It also allows for a clean
implementation of experimental acceptance cuts (i.e. detector
thresholds and/or detector geometry).  Hence, it should be
applicable to the description of the neutrino-pion production on
nuclei, in particular to the description of the
final-state-interactions (FSI) of the produced hadrons.

In our first investigation we have focused on resonant pion
production channels via the $\Delta_{33}(1232)$ resonance in charged
current (CC) reactions. The final-state-interactions  of the
resonance as well as the emitted pions have been calculated
explicitly for $^{12}C$ and $^{56}Fe$ nuclei and show a dominance of
pion suppression for pion momenta above 0.2 GeV/c due to $\Delta$
absorption on nucleons ($N\Delta \rightarrow NN$) as well as strong
pion rescattering in the kinematic regime of the $\Delta$ resonance
excitation. Neutrino reactions on $^{16}O$ are similar to $^{12}C$
targets and will be reported elsewhere. A comparison to integrated
$\pi^+$ spectra for $\nu_\mu + ^{12}C$ reactions demonstrates a
reasonable agreement in view of the uncertainties in the vector and
axial-vector form factors for the elementary reactions on nucleons.

We stress that precise double differential data (with respect to
$\nu$ and $q^2$) on free nucleons are mandatory before addressing the FSI
in a differential manner as proposed in this study. Once this task
is solved the question of in-medium effects on resonance transitions
in neutrino-nucleus reactions can be addressed and transition
amplitudes to higher baryonic resonances can be investigated. The
coupled-channel transport calculations used here already include
the various in-medium phenomena and provide a firm basis for the
analysis of future differential data.

\vspace{0.5cm} The authors like to acknowledge discussions with T.
Leitner who published a related study in the GiBUU model
\cite{Tinaneu} after submission of our manuscript and pointed out an
error in our earlier calculations.


\begin{thebibliography}{99}
\bibitem{xx1} E. A. Paschos, Nucl. Phys. B (Proc. Suppl.) 112 (2002) 89.
\bibitem{P1} E. A. Paschos, L. Pasquali and J. Y. Yu, Nucl. Phys. B
588 (2000) 263.
\bibitem{P2} E. A. Paschos and J. Y. Yu, Phys. Rev. D 65 (2002) 033002.
\bibitem{P3} S. L. Adler, S. Nussinov, and E. A. Paschos, Phys. Rev.
D 9 (1974) 2125.
\bibitem{NUANCE} H. Gallagher, D. Casper, Y. Hayato, and P. Sala,
Nucl. Phys. Proc. Suppl. 139 (2005) 278.
\bibitem{NEUT} G. P. Zeller, hep-ex/0312061; AIP Conf. Proc. 721
(2004) 375.
\bibitem{NEUGEN} Y. Hayato, Nucl. Phys. B (Proc. Suppl.) 112 (2002)
171.
\bibitem{NUX-FLUCA} D. Casper, Nucl. Phys.B (Proc. Suppl.) 112 (2002)
161.
\bibitem{Cass90} W. Cassing, V. Metag, U. Mosel, and K. Niita,
Phys. Rept. 188 (1990) 363.
\bibitem{Cass99}
  W.~Cassing and E.~L.~Bratkovskaya, Phys.~Rept.~{ 308} (1999)  65 .
\bibitem{Effe} M. Effenberger, E. Bratkovskaya and U. Mosel, Phys.
Rev. C 60 (1999) 044614; M. Effenberger and U. Mosel, Phys. Rev. C
62 (2000) 014605; P. M\"uhlich {\it et al.}, Phys. Rev. C 67
(2003) 024605.

\bibitem{Mueh} J. Lehr {\it et al.}, Nucl. Phys. A 671 (2000) 503;
Nucl. Phys. A 699 (2002) 324; Phys. Rev. C 68 (2003) 044603; Phys.
Rev. C 69 (2004) 024603.

\bibitem{Falter} T. Falter {\it et al.}, Phys. Lett. B 594 (2004) 61;
 Phys. Rev. C 70 (2004) 054609.
\bibitem{Segal} D. Rein and L. M. Sehgal, Ann. Phys. 133 (1981) 79.
\bibitem{Feynman} R. P. Feynman, M. Kislinger, and F. Ravndal, Phys.
Rev. D3 (1971) 2706.
\bibitem{r5} P. A. Schreiner and F. V. von Hippel, Nucl. Phys. B
58 (1973) 333.
\bibitem{Kolbe} G. Garvey, E. Kolbe, K. Langanke and S. Krewald,
Phys. Rev. C 48 (1993) 1919.

\bibitem{Wolf90} Gy. Wolf  {\it et al.}, Nucl. Phys. A
517 (1990) 615

\bibitem{Tina_dip} T. Leitner, {\it Diploma Thesis, University of Giessen
2005},\newline
http://theorie.physik.uni-giessen.de/documents/diplom/leitner.pdf

\bibitem{xxp} G. M. Radecky {\em et al.}, Phys. Rev. D34
(1986) 1161;
 J. Campbell {\em et al.}, Phys. Rev. Lett. 30 (1973) 335;
S. J. Barish {\em et al.}, Phys. Rev. D19 (1979) 2521; T. Kitagaki
{\em et al.}, Phys. Rev. D34 (1986) 2554.


\bibitem{URQMD1}
    S.A.~Bass {\it et al.},
    Prog. Part. Nucl. Phys. {42} (1998) 279.
\bibitem{URQMD2}
    M.~Bleicher {\it et al.},
    J. Phys. G {25} (1999) 1859.
\bibitem{Ehehalt}
  W.~Ehehalt and W.~Cassing, Nucl.~Phys. A 602 (1996)  449.
\bibitem{excita} W. Cassing, E. L. Bratkovskaya, and S. Juchem,
Nucl. Phys. A 674 (2000) 249.

\bibitem{Juchem} W. Cassing and S. Juchem, Nucl. Phys. A 665
(2000) 417; Nucl. Phys. A 672 (2000) 417.

\bibitem{Golubeva} W. Cassing, Ye. S. Golubeva, A. S. Iljinov, and
L. A. Kondratyuk, Phys. Lett. B 396 (1997) 26; Ye. S. Golubeva {\it
et al.}, Nucl. Phys. A 625 (1997) 832.

\bibitem{Mesh}  J. G. Messchendorp {\it et al.}, Eur. Phys. Jour. A11 (2001) 95

\bibitem{Wolf} Gy. Wolf, W. Cassing, and U. Mosel, Nucl. Phys. A 552 (1993)
549.
\bibitem{Peters} W. Peters, H. Lenske and U. Mosel, Nucl. Phys. A
640 (1998) 89.
\bibitem{Tina} T. Leitner, L. Alvarez-Ruso and U. Mosel,
{\it nucl-th/0511058}; L. Alvarez-Ruso, T. Leitner and U. Mosel,
{\it nucl-th/0601021}.


\bibitem{xxn} M.O. Wascko  {\em et al.}, in NuInt 05,
September 26 - 29, 2005, Okayama University, Okayama, Japan.
\bibitem{Tinaneu} T. Leitner, L. Alvarez-Ruso and U. Mosel,
{\it nucl-th/0601103}.
\end{thebibliography}
\end{document}